\documentclass[pra,showpacs,showkeys,twocolumn]{revtex4-1}
\usepackage[utf8]{inputenc}
\usepackage{amsmath}
\usepackage{amsfonts}
\usepackage{amssymb}
\usepackage{graphicx}
\newcommand{\xx}{\mathbf{x}}

\newcommand{\Meff}{M_{\rm eff}}

\usepackage{latexsym}
\usepackage{color}
\usepackage[dvipsnames]{xcolor}
\usepackage{textcomp}
\usepackage[normalem]{ulem}

\begin{document}
\title{Vortex pair annihilation in arrays of photon cavities}

\author{Vladimir N. Gladilin}
\author{Michiel Wouters}

\affiliation{TQC, Universiteit Antwerpen, Universiteitsplein 1,
B-2610 Antwerpen, Belgium}

\date{\today}

\begin{abstract}
We investigate theoretically the evolution of the vortex number in
an array of photon condensates that is brought from an incoherent
low density state to a coherent high density state by a sudden
change in the pumping laser intensity. We analyze how the
recombination of vortices and antivortices depends on the system
parameters such as the  coefficients for emission and absorption of
photons by the dye molecules, the rate of tunneling between the
cavities, the photon loss rate and the number of photons in the
condensate.
\end{abstract}

\maketitle

\section{Introduction}

Vortex excitations of superfluids~\cite{chaikin} play a crucial role
in a variety of phenomena such as the
Berezinskii-Kosterlitz-Thouless (BKT) transition, the Kibble-Zurek
(KZ) mechanism, the formation of Abrikosov vortex lattices and the
the drag force on objects moving through a superfluid. These
phenomena have been experimentally studied on a number of physical
platforms such as superconductors, liquid helium and ultracold
atoms~\cite{bec_book}. What all these systems have in common is that
they are up to a very good approximation in (local) thermal
equilibrium, a condition that is typically broken in experimental
platforms that are based on optical
systems~\cite{carusotto13,carusotto21}. Optical Bose-Einstein
condensates (BECs) have been experimentally achieved in optical
cavities filled with a dye molecule solution and in microcavities
where a photon is strongly coupled to an exciton resulting in
exciton-polariton~\cite{kavokin17}. While in the latter system,
interactions between exciton-polaritons can lead to stimulated
scattering into the ground state, in the photon-dye system it is
repeated absorption and reemission of photons \cite{klaers_therm}
that enables a macroscopic occupation of the ground
state~\cite{klaers10,marelic16,greveling18}.

Due to photon losses, optical BECs need constant pumping by an
excitation laser in order to reach a steady state where pumping
balances the losses. {Experimentally realized photon condensates are
therefore driven-dissipative systems. The availability of several
other experimental platforms, such as
exciton-polaritons~\cite{carusotto21}, superconducting circuits
\cite{carusotto20}  and Rydberg atoms \cite{bernien17}, for the
study of driven-dissipative systems and their interest for quantum
simulation and quantum computation has spurred substantial
theoretical activity
\cite{hartmann08,angelakis17,carusotto20,carusotto21}. }

A typical photon condensation experiment starts from an empty cavity
and optical excitations are created by turning on the pumping laser
where the phase transition is crossed when the photon density
exceeds threshold. Upon crossing the threshold, the system goes from
the disordered to the ordered state, which according to the KZ
mechanism may lead to the formation of vortex pairs. {Vortices in
nonequilibrium BECs have been the subject of both experimental
\cite{Kasprzak06,lagoudakis08,lagoudakis2011,Liew15,caputo2016} and
theoretical \cite{marchettiPRL10,cancellieriPRB14,ma2017,gladilin17}
study in exciton-polariton systems, whereas in nonequilibrium photon
condensates, vortices have only been studied
theoretically~\cite{gladilin20b}.}

{The vortex pairs that are generated after a pump power quench} are
expected to disappear in a subsequent phase annealing stage, such
that in the steady state above the critical density no vortices
remain. The recombination of vortices and antivortices depends on
their interactions, that has been shown to be crucially affected by
the nonequilibrium condition giving rise to particle currents away
from the vortex core. As a consequence, vortices, even of opposite
signs, experience a long-range repulsion~\cite{gladilin17}, that
obviously hampers their recombination. This physics was
theoretically investigated for exciton-polaritons and indeed it was
found that the vortex pair annihilation is strongly affected by the
nonequilibrium condition, where it was even observed that very long
lived complexes of vortices and antivortices may form under certain
conditions~\cite{gladilin19}.

Photon condensates are even somewhat more peculiar for the study of
vortex physics, because their negligible interactions imply that the
core size tends to infinity in the equilibrium limit. This stands in
contrast to polariton condensates, where the interactions always
keep the vortex core size finite. It was however shown recently that
a finite vortex core and therefore well defined vortex excitations
do exist in lattices of coupled nonequilibrium photon
condensates~\cite{gladilin20b}. Such lattices of photon condensates
can be experimentally created by patterning the microcavity mirrors
\cite{kassenberg20,dung17}, analogous to lattices for
exciton-polaritons \cite{schneider16}. These vortices feature
several unusual properties such as self acceleration and core
deformation in addition to the aforementioned outward currents.

{The twodimensional nature of polariton condensates implies that the
phase transition between the normal and superfluid state} at
equilibrium is of the BKT type. Numerical simulations have provided
evidence that a similar transition exists out of equilibrium
\cite{dagdavorj,caputo2016,gladilin19,comaron18}, even though
studies based  on the renormalization group have pointed out that
the phase transition is crucially affected at long distances by the
Kardar-Parisi-Zhang dynamics of the phase, that tends to destroy the
ordered phase~\cite{altman15,wachtel,sieberer,zamora}. In practice,
however, the KPZ physics can be limited to very large system sizes,
so that in experimental 2D systems the BKT-like physics dominates
\cite{dagdavorj,caputo2016,gladilin19,comaron18}. Recently, we have
shown by analytical arguments and numerical simulations that also
photon condensates without interactions do feature a BKT like phase
transition, that is stabilized by the driving and
dissipation~\cite{gladilin21}. The question of the dynamical
formation of the ordered state after a rapid switching of the
pumping intensity is therefore also relevant for photon condensates.

In the present paper, we address this issue. In particular, we
investigate the annihilation of vortex pairs in a lattice of photon
condensates where the pumping laser intensity is suddenly ramped up,
starting from zero.  After a short initial transient, the density
goes relatively quickly to its steady state, but many phase defects,
remnants of the random phases of the initial state, remain. By
varying the system parameters (tunneling strength,
emission/absorption coefficient, loss rate, photon number) around
the range of their typical experimental values, we try to shed more
light on the physics that governs the annihilation dynamics. We
reveal a regime where the quasi-circular vortex trajectories
together with the long range repulsion lead to a low mobility
vortex-antivortex phase that shows very slow annihilation. In
addition, we show that for very high emission/absorption rate the
annihilation is fast and tends to the time dependence derived for
the relaxational dynamics of the XY model~\cite{yurke1993,jelic11},
a regime that was previously also obtained in numerical simulations
for polariton condensates close to thermal equilibrium
\cite{comaron18,gladilin19}.

{The remainder of the paper is organized as follows. In Sec.
\ref{sec:model}, we recapitulate our classical field model for
photon condensates and describe the numerical simulations. In Sec.
\ref{sec:disc}, we discuss the various dependencies of the pair
annihilation rate on the system parameters: the rates of
emission/absorption, tunneling and losses and the photon number. Our
conclusions are presented in Sec. \ref{sec:concl}.}

\section{Model \label{sec:model}}

\begin{figure} \centering
\includegraphics[width=0.8\linewidth]{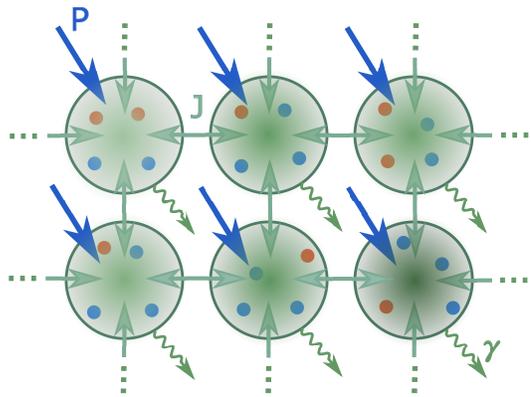}
\caption{Sketch of a part of the photon condensate array under
study: the photonic cavity modes are coupled through tunneling with
amplitude $J$ and are lost from the cavities at rate $\gamma$. The
dye molecules are localized in the cavities and can be in the ground
state or excited state manifolds (indicated by different colors).
The system is pumped by a laser with intensity $P$, that excites the
dye molecules.} \label{fg0}
\end{figure}

{ At the semiclassical level, the system of coupled photon
condensates can be described by a generalized Gross-Pitaevskii
equation for the photon field $\psi(\xx)$ at each cavity position
$\xx$ ~\cite{gladilin20}
\begin{align}
i \hbar \frac{\partial \psi(\xx,t)}{\partial t} = &
\frac{i}{2}\left[ B_{21}M_2(\xx,t) - B_{12}
M_1(\xx,t)-\gamma\right]\psi(\xx,t) \nonumber \\
&- (1-i\kappa) J  \sum_{\xx' \in \mathcal N_\xx} \psi(\xx',t),
 \label{eq:gGP}
\end{align}
where $\gamma$ is the photon loss rate and $J$ the coupling between
the nearest-neighbor cavities \cite{kassenberg20,dung17}. The
photons thermalize thanks to repeated absorption and emission by the
dye molecules with respective rate coefficients $B_{12}$ and
$B_{21}$. The ground (excited) molecular state occupation is denoted
by $M_{1(2)}$ and obeys $M_1(\xx)+M_2(\xx)=M$, where $M$ is the
number of dye molecules at each lattice site. The emission and
absorption coefficients of the thermalized molecules satisfy the the
Kennard-Stepanov relation \cite{kennard,stepanov,moroshkin14}
$B_{12} = e^{\beta \Delta} B_{21}$ where $\Delta$ is the detuning
between the cavity frequency and the dye zero-phonon transition
frequency.} { The Kennard-Stepanov relation for the absorption and
emission coefficients leads to energy relaxation with dimensionless
strength $\kappa= B_{12} \bar M_1/(2T)$ (we set the Boltzmann
constant $k_B=1$) \cite{gladilin20}. }

{The noise enters in the model due to fluctuations stemming from
spontaneous emission that are described by adding a unit modulus
complex number with random phase to the photon amplitude at the
spontaneous emission rate $B_{21} M_2$~\cite{henry82,verstraelen19}.
We have checked numerically that this way of introducing
stochasticity yields the same results as a Langevin noise term
representing the shot noise in the emission and absorption
processes. }

{ Equation \eqref{eq:gGP} is coupled to a rate equation describing
the evolution of the number of excited molecules due to emission and
absorption processes and external pumping. The latter has to be
applied to compensate for the photon losses. The steady-state
average value $\bar n$ of the photon number in each cavity, $n =
|\psi(\xx)|^2$, results from the balance between the pumping rate
$P$ and losses and satisfies $P = \gamma \bar n$. } {Under the
condition $J \ll T$, which assures that the occupations of all
momentum states are much larger than one, the generalized Gross
Pitaevskii classical field model \eqref{eq:gGP} is valid for all the
modes and there is no need to use a more refined quantum optical
approach \cite{kirton13,kirton15,kirton16}.}

The noise, inherent to the spontaneous emission by the dye
molecules, leads to the density and phase fluctuations. For the
simplest case of a single cavity, a crossover in the density
fluctuations between a `grandcanonical' regime with large
fluctuations ($\delta n^2 \sim \bar n^2$), for $\bar n^2\ll M_{\rm
eff}$, and a `canonical' regime with small fluctuations ($\delta n^2
\ll \bar n^2$), for $\bar n^2\ll M_{\rm eff}$, have been observed
\cite{klaers12,schmitt14}. Here, the `effective' number of molecules
is given by $\Meff = (M+\gamma e^{-
\Delta/T}/B_{21})/[2+2\cosh(\Delta/T)]$. For $e^{\Delta/T}\ll 1$,
one has $M_{\rm eff} \approx \bar M_2 \approx \eta M e^{\Delta/T}$
with $\eta=1+{\gamma }/(2 \kappa T)$, while the energy relaxation
parameter becomes $\kappa\approx B_{21}Me^{\Delta/T}/(2T)$.

In our simulations, the following parameters are fixed: the
temperature $T=25$ meV, the energy detuning $\Delta=-180$ meV, and
the total number of molecules $M=5\times 10^{10}$. With these dye
molecule parameters, $M_{\rm eff}\approx 3.73\times 10^7$ and
$n^2/M_{\rm eff}$ varies from 67 for $n=n_0$ down to 0.026 for
$n=0.02 n_0$, where $n_0=5\times 10^4$ is our unit for the photon
density. The other relevant parameters are expressed in the
following units: $B_0= 10^{-7}$ meV for the emission coefficient,
$J_0=0.02$ meV for the coupling strength, and $\gamma_0=0.02$ meV
for the loss rate. In each simulation, the initial state is formed
by random values of the phase and (very small) number of photons
($n< 10^{-3}\bar n$) in each cavity, with no correlation between
cavities. The pair annihilation dynamics are studied for an array of
$200\times 200$ coupled photon cavities with periodic boundary
conditions. Sample-averaged curves correspond to 10 to 32 runs.

\section{Discussion of the numerical results \label{sec:disc}}

\subsection{The role of the emission-absorption rates}

Photon condensates reach thermal equilibrium in the steady state
when the loss rate is much smaller than the rates of absorption and
re-emission. Even though in the present study, we do not investigate
the steady state, one can still expect that the relative importance
of emission-absorption with respect to losses has some influence on
the phase relaxation dynamics. In our simulations, we have first
kept the loss rate fixed at $\gamma_0=0.02$ meV and varied the
emission coefficient $B_{21}$ and correspondingly the absorption
coefficient according to $B_{12} = e^{\beta \Delta} B_{21} $ at
fixed detuning.

\begin{figure} \centering
\includegraphics[width=1.\linewidth]{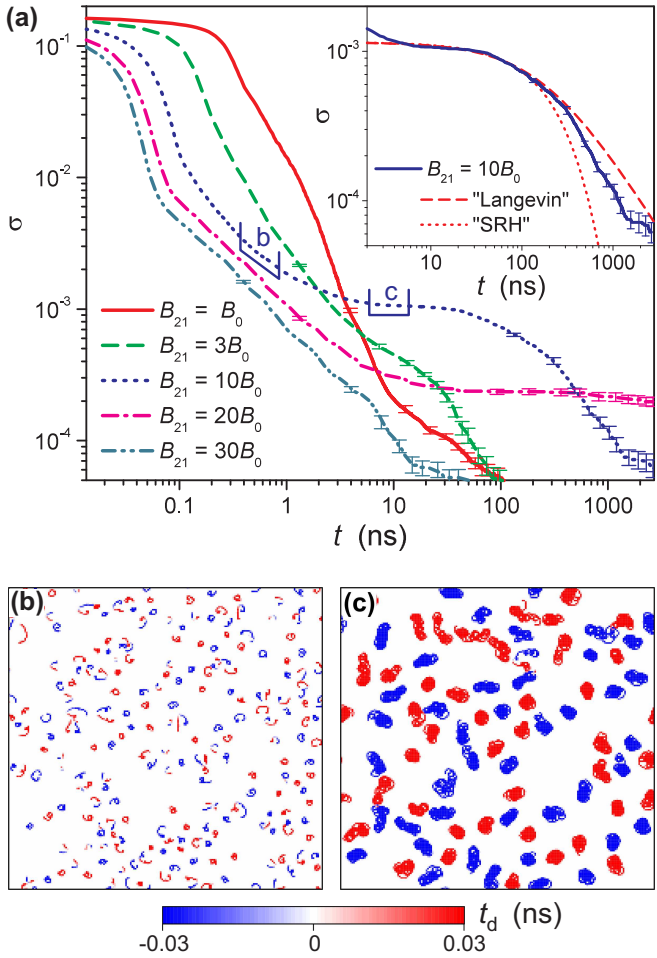}
\caption{(a) Evolution of the vortex-pair density $\sigma$ at
different values of the emission coefficient $B_{21}$ and fixed
$J=J_0$, $\gamma=\gamma_0$, $\bar n=n_0$. Inset: long-time dynamics
of $\sigma$ at $B_{21}=10B_0$ (solid line) in comparison with the
``Langevin'' (dashed line $\sigma=1.15\times
10^{-3}\left[1+{t}/(189\ \rm{ns})\right]^{-1}$) and ``SRH'' (dotted
line $\sigma=1.15\times 10^{-3}\exp\left[-{t}/(220\
\rm{ns})\right]$) recombination dynamics. Panels (b) and (c) show
trajectories of vortices (red) and antivortices (blue) in the array
with $B_{21}=10B_0$ on the time intervals labeled in panel (a) with
``b'' and ``c'', respectively. $t_{\rm d}$ is the time the core
dwells in a given unit cell of the array, with the sign of the
dwelling time representing the vorticity sign.} \label{fg1}
\end{figure}

Figure~\ref{fg1}(a) shows the time evolution of the vortex-pair
density $\sigma$ (the number of pairs per cavity) for various
emission coefficients $B_{21}$ on a double logarithmic scale. Quite
remarkably, we see a rich behavior in the evolution of the vortex
pair density. Starting with the lowest value of $B_{21}$ (red full
line), we see after an initial constant at very short times a fast
decay up to $t\approx 10$ ns and a subsequently slower decay up to
$t=100$ ns, when only two vortex pairs remain in the simulation
region. {The relatively slow time scale on which the vortex pair
annihilation manifests itself is promising for its experimental
detection: the observation of dynamics on the ns scale or slower is
well within the domain of standard electronics.  }

The initial pair number annihilation rate increases when the
emission coefficient $B_{12}$ is increased. There are two mechanisms
that explain this dependence: (i)  an increase of $B_{21}$ implies a
more efficient interaction of cavity photons with the reservoir
formed by dye molecules, and hence a more efficient role of pumping
that tends to keep the condensate density close to its steady-state
value $\bar n$~\cite{gladilin19,gladilin19b}. Larger $B_{21}$
therefore more strongly disfavors states with many vortex cores,
since in these states there are many cavities with photon densities
 much smaller than
$\bar n$. (ii) An increase in $B_{21}$ implies an increase in
$\kappa \approx B_{21}Me^{\Delta/T}/(2T)$. According to Eq.
\eqref{eq:gGP}, the effect of $\kappa$ is to eliminate the photon
field's phase and density gradients.  A larger value of $\kappa$
therefore leads to a faster recombination of vortices and
antivortices.

\begin{figure} \centering
\includegraphics[width=1.\linewidth]{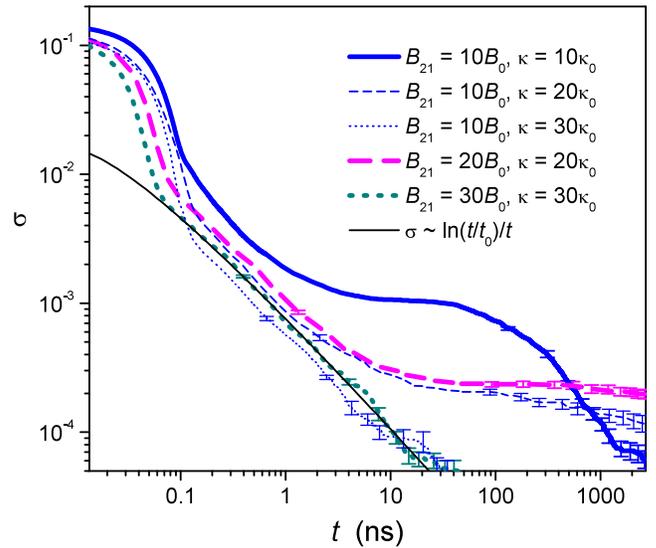}
\caption{Evolution of the vortex-pair density $\sigma$ at different
values of the emission coefficient $B_{21}$ and damping parameter
$\kappa$, expressed in units of $\kappa_0=0.076$, in the case, where
these two parameters are treated as independent of each other, and
for fixed $J=J_0$, $\gamma=\gamma_0$, $\bar n=n_0$. The thin solid
line corresponds to a decay $\sigma=0.043(t/t_0)^{-1}\ln (t/t_0)$
with $t_0=3$ ps.} \label{fg2}
\end{figure}

Figure \ref{fg2} illustrates that, at $\kappa>1$, it is mainly the
value of $\kappa$ that determines the vortex-pair annihilation rate.
It shows the time evolution of the pair number for different values
of $B_{21}$ with an independent variation of $\kappa$ (i.e. not
according to $\kappa = B_{12} \bar M_1 /(2T)$). Curves with the same
$\kappa$ but with different $B_{21}$ are close to each other,
clearly illustrating that the pair number dynamics is mainly
governed by $\kappa$. The thin solid line in Fig. \ref{fg2} shows
the analytical dependence for the number of vortices obtained for
relaxation dynamics of the two-dimensional XY model
$\sigma(t)\propto (t/t_0)^{-1}\ln (t/t_0)$ ~\cite{yurke1993,jelic11}
and agrees well with the numerics for the largest values of
$\kappa$. This agreement is in analogy with simulations done for
exciton-polariton condensates ~\cite{comaron18,gladilin19}. Also in
line with the current observations, it was found in the polariton
case that the vortex annihilation is accelerated when increasing
$\kappa$ \cite{gladilin19}.

Going back to Fig. \ref{fg1}, one observes that the long-time
behavior of $\sigma(t)$ varies non-monotonously with $B_{21}$. For
$B_{12}=10 B_0$, a clear plateau (on the used log-log scale)
develops between $t=10$ ns and $t=100$ ns. We propose as the
mechanism for the formation of the plateau the interplay between
vortex-antivortex distances and the radius of the quasi-circular
trajectories of self-accelerated vortices and antivortices, which
decreases with increasing $B_{21}$~\cite{gladilin20b}.
{Quasi-circular vortex trajectories with relatively small radius
constitute an important distinctive feature of the system under
consideration as compared to the exciton-polariton condensates
analyzed in Refs.~\cite{gladilin17,gladilin19}.} The quasi-circular
form of the trajectories leads to a localization of vortices,
hampering their encounters when the vortex density becomes low. This
is illustrated in panels (b) and (c) of Fig. \ref{fg1}, with their
time intervals indicated in panel (a) as ``b'' and ``c''
respectively. In panel (b), with the higher vortex density,
trajectories of vortices (red) and antivortices (blue) intersect
quite frequently, leading to efficient annihilation. Within the
corresponding subnanosecond time interval, as many as 70
vortex-antivortex pairs annihilate in the simulation region. In
contrast, the low vortex density in panel (c) together with the well
localized vortex orbits prevents annihilation during the longer time
interval ``c''. In addition to the self-acceleration that leads to
circular orbits, the nonequilibrium situation leads to outward
particle flows from vortex cores that cause {- similarly to the case
of exciton-polariton condensates~\cite{gladilin17} -} long-range
repulsion of vortices independent of their mutual
chirality~\cite{gladilin20b}. As a consequence, the localized orbits
show very little overlap in the regime of low density. Furthermore,
most of these orbits are seen to have a very low mobility, similar
to a reduced transverse mobility of electrons in a magnetic field,
so that vortex-antivortex collisions are rather rare. As a result,
only 1 pair recombines in the simulation region in the time interval
of about 13 ns. Increasing the emission rate $B_{21}$ both reduces
the radius {of the circular vortex trajectories} and increases the
long range repulsion \cite{gladilin20b}, explaining why the plateau
only develops for sufficiently large $B_{21}$.

The subsequent decay of the pair density for $B_{12}=10B_0$ can be
understood from the recombination of low mobility particles,
analogous to particle-hole recombination in low mobility
semiconductors. In the inset of Fig. \ref{fg1}(a), we compare
$\sigma(t)$ with fits to the Langevin pairwise and
Shockley-Read-Hall (SRH) single particle recombination, described by
\begin{align}
\frac{d}{dt} n_p &= - \alpha\; n_p^2 \hspace{0.5cm} {\rm (Langevin)}, \\
\frac{d}{dt} n_p &= - \alpha\; n_p \hspace{0.5cm} {\rm (SRH)}.
\end{align}
The better fit to the Langevin model for times up to about 500 ns
shows that pairwise annihilation is dominant when the pair density
is sufficiently large. At later times, when the pair density has
dropped, there is some deviation towards the SRH. The vortex density
evolution is seen to be relatively close to the Langevin dynamics
but with some deviation towards the SRH. Our tentative explanation
of this deviation is that -- as implied by Fig. \ref{fg1}(c) -- not
all vortices have the same mobility. Indeed, the exact SRH
dependence would correspond to the limiting case where most of the
vortices are pinned (have zero mobility) and some are mobile. Then
it is natural to expect that less drastic but still significant
differences in mobility result in some ``intermediate'' annihilation
curve as compared to the purely Langevin and SRH dynamics.

As seen from Fig.~\ref{fg1}(a), when increasing the emission
coefficient from $B_{21}=10B_0$ to $B_{21}=20B_0$, the plateau on
the $\sigma(t)$ shifts down to lower pair densities and it fully
disappears at $B_{21}=30B_0$. At those large $B_{21}$ the energy
relaxation parameter $\kappa$ exceeds 1 and the effect of its
increase on the long-time pair-density dynamics dominates over that
of the emission-absorption rates.

At the longest times, when only a few pairs remain ($\sigma \lesssim
10^{-4}$), the slope of $\sigma(t)$ becomes flatter. In this regime,
there is a formation of metastable low-density vortex-antivortex
configurations. This interpretation is supported by the results of
our simulations for finite arrays, where formation of metastable
vortex-antivortex states, qualitatively resembling those obtained in
gGPE simulations for exciton-polariton
condensates~\cite{gladilin19}, has been observed \cite{suppl}.

\subsection{The role of tunneling and losses}

\begin{figure} \centering
\includegraphics[width=1.\linewidth]{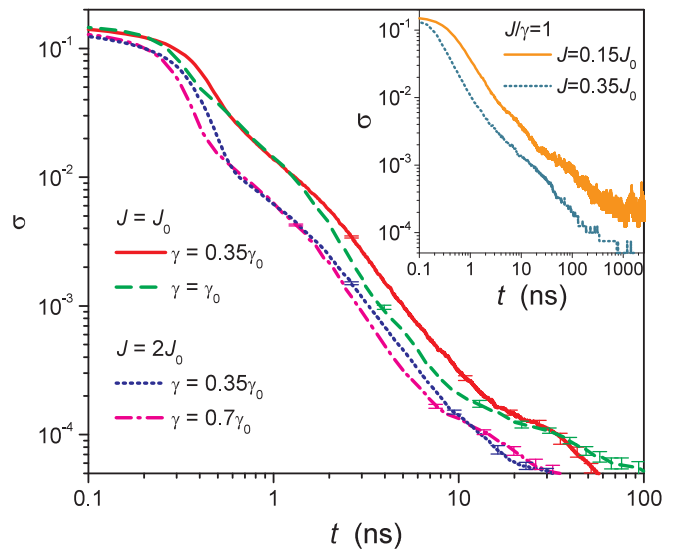}
\caption{Evolution of the vortex-pair density $\sigma$ at different
values of the coupling strength $J$ and loss rate $\gamma$ at fixed
$B_{21}=B_0$ and $\bar n=n_0$. Inset: single-run evolution of
$\sigma$ at two different relatively small values of $J$ and fixed
$J/\gamma=1$, $B_{21}=3B_0$, $\bar n=n_0$.} \label{fg4}
\end{figure}

From the two parameters $J$ and $\gamma$ out of which the length
scale $r_v\propto \sqrt{J/\gamma}$ has been argued to serve as a
first estimate for the vortex core size~\cite{gladilin20b}, the
effect of the tunneling on the pair relaxation dynamics is the most
straightforward. Comparing in Fig. \ref{fg4} the solid to the dotted
curve, one sees that the pair relaxation rate is enhanced when $J$
is increased, a dependence that can be understood with the following
``spatial resolution'' argument. An increase of $J$ can be thought
of as an increase in spatial resolution within the continuum limit
of the lattice model~\cite{gladilin20b}. Correspondingly, the
long-time behavior of $\sigma(t)$ at $J=2J_0$ is close to that of
$\sigma(t)/2$ at $J=J_0$, except at shortest $t$, when the vortex
cores strongly overlap and the structure of individual vortices is
not well developed, so that the spatial resolution argument is not
applicable.

In contrast to the dependence on $J$, the effect of the loss rate
$\gamma$ is more subtle. The comparison of the solid and dashed
lines in Fig. \ref{fg4} shows that for larger $\gamma$, relaxation
is faster at intermediate times, but slower at long times. Our
interpretation of these dependencies is as follows. At earlier
times, when the vortex density is large, it is mainly the enhanced
stabilization of the density due to pump-loss dynamics that leads to
a faster disappearance of vortex pairs. This argument is in line
with the explanation of the role of $B_{21}$ on the very early time
dynamics in the discussion of Fig. \ref{fg1}. At lower pair density
however, the outward particle currents, leading to repulsion between
vortices of opposite chirality, become important~\cite{gladilin20b}.
These currents increase with the loss rate and therefore stronger
losses suppress the pair annihilation rate. In general, the
competition between the two different effects of losses on the
annihilation rate results in a rather intricate influence of
$\gamma$ on the shape of the annihilation curve. When $J$ and
$\gamma$ are varied simultaneously, keeping the length scale
$r_v=\sqrt{J/\gamma}$ fixed, the effect of $J$ on $\sigma(t)$
dominates, as illustrated in the inset of Fig. \ref{fg4}: the curve
with larger $J$ (dotted) shows a faster decrease of $\sigma$ with
respect to the one at lower $J$ (full line).

A remarkable feature of the curves $\sigma(t)$, shown in the inset
of Fig.~\ref{fg4} and corresponding to {rather} weak coupling $J$,
is their non-monotonous behavior at relatively low pair densities.
This behavior reflects noise-induced generation of new vortex pairs,
which becomes more pronounced with decreasing $J$ as the system
gradually approaches the BKT transition point~\cite{gladilin21}.
These pair generation processes significantly slow down relaxation
of the system towards a vortex-free state.

\subsection{The influence of the photon number}

\begin{figure} \centering
\includegraphics[width=1.\linewidth]{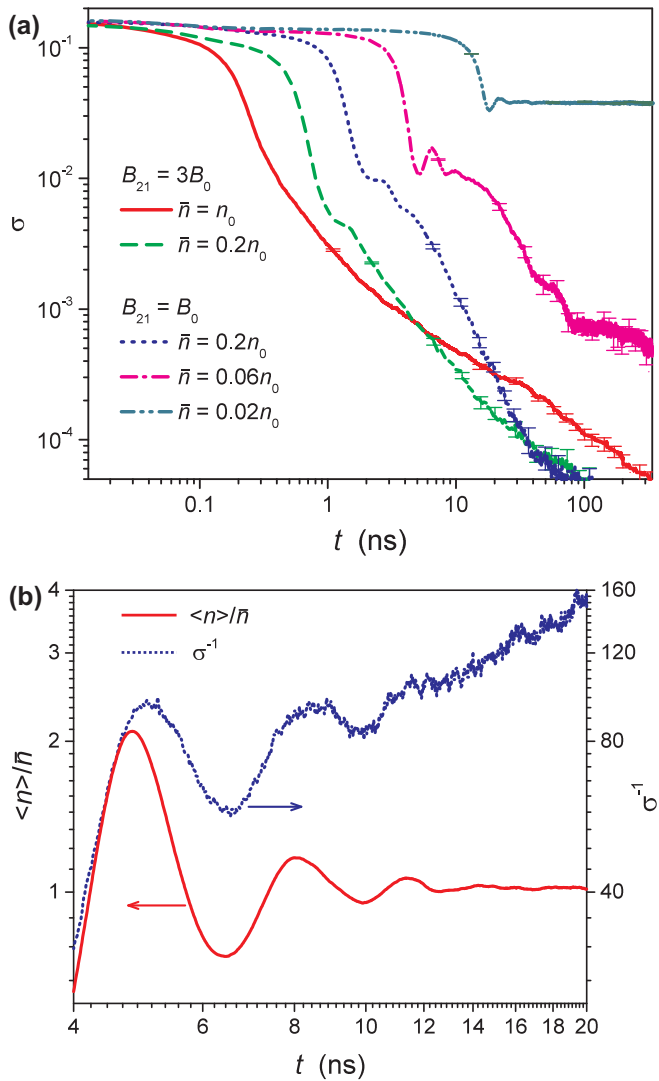}
\caption{(a) Evolution of the vortex-pair density $\sigma$ at
different values of the emission coefficient $B_{21}$ and average
photon density $\bar n$ for fixed $J=J_0$ and $\gamma=0.35\gamma_0$.
(b) Oscillations of the inverse vortex-pair density $\sigma^{-1}$
(dotted line) in comparison to the oscillations of the spatially
averaged photon density $\langle n \rangle$ (solid line) at
$B_{21}=B_0$, $\bar n=0.06n_0$.} \label{fg5}
\end{figure}

Let us finally discuss how the photon number affects the pair
annihilation rate. In Fig. \ref{fg5}(a), we show $\sigma(t)$ for
various photon numbers, that are smaller than the ones used in Figs.
\ref{fg1} - \ref{fg4}.

The first effect of a decreasing photon number is that it takes
longer before the appreciably fast pair annihilation starts. This is
the same trend as in Fig.~\ref{fg1} for decreasing emission rate.
The reason is analogously that for lower photon number, it takes
longer for the density to stabilize around its mean value, because
of slower absorption-emission dynamics.

A second feature that stands out in Fig. \ref{fg5}(a) is that at the
smallest photon number $n=0.02 n_0=1000$, the pair density does not
decay to zero (dash-dot-dotted line). The reason is that the photon
number is below the critical photon number for the
Berezinskii-Kosterlitz-Thouless phase transition $n_c\approx 1200$,
estimated with the use of the numerical and analytical results from
Ref. \cite{gladilin21}. The vortices present at late times (after
$t\approx 30$ ns) are therefore spontaneous vortices caused by the
noise in the system and are no longer due to the random initial
condition of the photon field. For the dash-dotted curve at a the
second lowest photon number $n=0.06n_0=3000$, creation of new pairs
by the noise leads to pronounced fluctuations of the pair density
and strongly slows down its overall final decay. Since the photon
number is much above the critical number for the BKT transition, we
expect that the pair number will tend to zero at very long times.

A further notable feature of the two curves at the lowest photon
numbers are the oscillations after the initial decay of $\sigma$.
These are related to the behavior of the photon number density [see
Fig. \ref{fg5}(b)], that exhibits relaxation oscillations analogous
to lasers~\cite{svelto} or perturbed photon condensates
\cite{ozturk19}. At lower densities, the noise becomes relatively
more important, leading to a more intense creation of new vortex
pairs. Consequently, the vortex number shows oscillations, which
decay together with the photon density oscillations and have the
opposite phase (with some retardation).

Let us now turn to the difference between the solid and dashed lines
in Fig. \ref{fg5}(a), that illustrate most clearly that far above
the BKT transition, a lower photon number speeds up the decay of
vortex pairs. We identify two reasons for this dependence. First, at
lower photon number, the outward currents from each vortex core,
which are responsible for the vortex-antivortex repulsion, decrease
\cite{suppl}, facilitating the annihilation of vortex pairs.
Secondly, at lower photon numbers, the larger radius of the
quasi-circular vortex trajectories and higher relative intensity of
the noise \cite{suppl} give the vortices a higher mobility, such
that the formation of quasi-stable vortex patterns similar to that
shown in Fig. \ref{fg1}(c) is less likely.

\section{Conclusions and outlook \label{sec:concl}}
We have studied in this paper the phase annealing of a lattice of
nonequilibrium photon condensates, that is created by a rapid
switching of the pump laser intensity. The initial state shows very
poor coherence, due to its random phases at each lattice site. After
a transient with large density fluctuations, well defined vortices
and antivortices are formed.  The energy relaxation, that stems in
the case of photon condensates from the Kennard-Stepanov relation,
leads subsequently to the annihilation of vortex pairs. Our
numerical simulations show that when the emission and absorption
rates are very high, the vortices annihilate with the time
dependence from XY model relaxational dynamics
~\cite{yurke1993,jelic11}.

At lower values of the emission/absorption rates, vortex repulsion,
caused by outward currents from the vortex cores, in combination
with quasi-circular motion of self-accelerated vortices can
considerably slow down the pair annihilation. In this regime, a
repulsive gas of vortices and antivortices with low mobility is
formed, that survives for a significantly longer time. { Its
annihilation was shown to be in between Langevin and SRH-like
dynamics.

{For what concerns the influence of the tunneling and loss rates,
the dependence on the former is the most straightforward. A
modification of the tunneling rate can be seen as a change in
spatial resolution with a corresponding effect on the vortex density
(an increased tunneling leads to a decrease in the number of
vortices). The dependence of the vortex annihilation on the loss
rate is less pronounced and exhibits different trends at early and
late times. Increasing the losses leads to an increase of the
annihilation rate at early times and a slowing down at later times.
}

{Finally, we have investigated the role of the photon number.
Naturally, when the photon number is decreased below the critical
number for the BKT transition, the vortex pairs no longer disappear
at late times, but an equilibrium between their recombination and
their noise-induced generation is established. Within the ordered
phase however, increasing the photon number slows down the vortex
pair annihilation due to an increase of outward currents and a
reduction of the vortex mobility. }

{In the present work, we have focused on the presence/absence of
vortices, the natural perspective from the phase annealing of 2D
equilibrium condensates. It has been shown however that for
sufficiently large systems, the phase dynamics of nonequilibrium
condensates belongs to the KPZ universality class. It would be
interesting to look for signatures of the KPZ physics on the phase
annealing of nonequilibrium photon condensates.}

{Our simulations were performed for a translationally invariant
system with equal tunnelings between all lattice sites, where the
ground state is one with a uniform phase. The time needed to
eliminate the vortices from the system is therefore the time that
the system needs to reach the lowest energy state. It has been
suggested to use photon condensates for analog computation by
letting them find the ground state of a nontrivial classical XY
model hamiltonian~\cite{kassenberg20}. Our insights obtained here on
the annihilation of vortex pairs are expected to be relevant for the
estimation of the time needed to reach the ground state in such
applications and the trends that were observed for the dependence on
the system parameters may be useful in order to improve the
performance of analog computing with photon BECs.}

\section*{Acknowledgements}

VG was financially supported by the FWO-Vlaanderen through grant nr.
G061820N.

\end{document}